\documentstyle[epsf,emulateapj,pstricks]{article}

\def\einstein	{{\em Einstein}\/}
\def\exosat     {{\em EXOSAT}\/}
\def\asca	{{\em ASCA}\/}
\def\ginga	{{\em Ginga}\/}
\def\rosat	{{\em ROSAT}\/}
\def\bohringer	{B\"{o}hringer}
\def\am		{$^\prime$}
\def\deg	{$^{\circ}$}
\def\hfifty     {$H_0$=50~km$\;$s$^{-1}\,$Mpc$^{-1}$}

\begin{document}

\lefthead{MARKEVITCH \& VIKHLININ}
\righthead{\asca\ AND \rosat\ CLUSTER TEMPERATURES}

\title{COMPARISON OF \asca\ AND \rosat\ CLUSTER TEMPERATURES: A2256, A3558
AND AWM7}

\author{Maxim Markevitch\altaffilmark{1}}
\affil{University of Virginia, Astronomy Department, Charlottesville, VA
22903; mlm5y@virginia.edu}

\and

\author{Alexey Vikhlinin\altaffilmark{1}}
\affil{Harvard-Smithsonian Center for Astrophysics, 60 Garden St.,
Cambridge, MA 02138; alexey@head-cfa.harvard.edu}

\altaffiltext{1}{Also IKI, Profsoyuznaya 84/32, Moscow 117810, Russia}

\centerline{\small\em Accepted for ApJ, 1996 July 24}

\begin{abstract}

We address the consistency between \asca\ and \rosat\ spatially-resolved
cluster temperature measurements, which is of significant immediate interest
given the recent \asca\ reports of temperature gradients in several hot
clusters. We reanalyze \rosat\ PSPC data on A2256 (originally analyzed by
Briel \& Henry) using the newer calibration and a technique less sensitive
to the calibration uncertainties. We find a temperature decline with radius
in good agreement with \asca's, although with much larger errors than those
of \asca. We also present \asca\ temperature maps and radial profiles of two
cooler clusters, A3558 and AWM7. These are compared to the \rosat\ results
of Bardelli et al.\ and Neumann \& \bohringer.  In A3558, our analysis
detects an asymmetric temperature pattern and a slight radial
temperature decline at $r\sim 0.5\,h^{-1}$ Mpc, in addition to the spectral
complexity of the cD galaxy region. We do not detect any significant spatial
temperature variations in AWM7, except around the cD galaxy, in agreement
with the earlier \asca\ analysis of this cluster. Radial temperature
profiles of these two clusters are in a qualitative agreement with \rosat.
However, while \asca\ average temperatures (5.5 and 3.9~keV, respectively)
agree with those from other high-energy instruments (\ginga\ and \exosat\
for A3558 and \einstein\ MPC and \exosat\ for AWM7), \rosat\ temperatures
are lower by factors of 1.7 and 1.25, respectively. We find that including
realistic estimates of the current \rosat\ systematic uncertainties enlarges
the temperature confidence intervals so that \rosat\ measurements are
consistent with others for these two clusters as well. Due to the limited
energy coverage of \rosat\ PSPC, its results for the hotter clusters are
highly sensitive to calibration uncertainties. We conclude that within the
present calibration accuracy, there is no disagreement between \asca\ and
other instruments. This adds confidence to the \asca\ results on the hotter
clusters which can at present be studied with precision only with this
instrument. On the scientific side, a \rosat\ temperature underestimate for
A3558 may be responsible for the anomalously high gas to total mass fraction
found by Bardelli et al.\ in this core member of the Shapley supercluster.

\end{abstract}


\section{INTRODUCTION}

In this paper, we focus on a comparison of \asca\ and \rosat\ spatially
resolved temperature measurements of clusters of galaxies. Such measurements
are of great importance; for example, they are needed for the determination
of cluster masses, a problem which has fundamental cosmological implications
(e.g., White et al.\ 1993). Resolved temperature measurements for a
significant number of hot (and therefore most massive) clusters have only
become possible now with the advent of \asca\ with its imaging capability
and the 0.5--11 keV energy coverage (Tanaka et al.\ 1994). Recent \asca\
papers on several hot clusters have reported the detection of unexpectedly
strong temperature gradients in the outer cluster regions in several cases
(e.g., in A2163, A665, A2256, A2319, Markevitch et al.\ 1996; Markevitch
1996, hereafter M96; A2218, Loewenstein et al.\ 1996). These observations
contradict the common assumption that clusters are isothermal. \asca\
analysis techniques are relatively complex due to the properties of the
\asca\ mirrors and at present are far from being established. Therefore, it
is necessary to independently confirm the \asca\ findings for at least those
clusters for which other data are available. \rosat\ PSPC data with good
spatial resolution exist for many clusters. For the hotter objects, however,
these data are only of limited use, due to the weak temperature dependence
of the cluster spectrum shape within the PSPC 0.1--2.5~keV energy band,
which means that any calibration uncertainty translates into a large error
in the measured temperature, even for observations with good statistics.
Nevertheless, some temperature estimates can be derived with \rosat. Below
we compare \asca\ and \rosat\ results on three clusters with different
temperatures and either absent or weak cooling flows (to avoid complication
of the \asca\ analysis), for which there are published \rosat\ results and
publicly available \asca\ data.  These are the hot cluster A2256
($z=0.058$), cooler, rich cluster A3558 ($z=0.048$) and the cool nearby
group AWM7 ($z=0.018$). \asca\ results on A2256 have been reported in M96;
Churazov et al.\ (1996) obtained generally similar results using an
independent technique. \rosat\ data on A2256 have been previously analyzed
by Briel \& Henry (1994, hereafter BH) who found a temperature profile
different from (although not inconsistent with) M96. BH also found two
high-temperature regions in this cluster, whose existence was not confirmed
by \asca\ (M96). \rosat\ results on A3558 have been reported by Bardelli et
al.\ (1996, hereafter B96), while those on AWM7 were given by Neumann \&
\bohringer\ (1995, hereafter NB).  Full analysis of the AWM7
\asca\ observations was presented by Ohashi et al.\ (1994) and Ezawa et al.\
(1996), as well as by Loewenstein (1995). In this paper, we perform a
detailed reanalysis of A2256 \rosat\ PSPC data and discuss the sources of
the apparent discrepancy with the results of BH. We also present the first
analysis of \asca\ data on A3558 and our independent reanalysis of the
\asca\ central pointing toward AWM7, as well as a brief reanalysis of
\rosat\ A3558 and AWM7 observations for which we confirm the results
published earlier. In this paper, we discuss the values of the gas
temperature projected onto the image plane (as opposed to those for
three-dimensional cluster regions), to facilitate comparison with other
work. A value of \hfifty\ is used throughout the paper.

\begin{figure*}[ht]
\small
\renewcommand{\arraystretch}{1.3}
\renewcommand{\tabcolsep}{3.5mm}
\begin{center}
\vspace{-3mm}
TABLE 1
\vspace{2mm}

{\sc Results of fitting A2256 \rosat\ PSPC data using different methods}
\vspace{2mm}

\begin{tabular}{lccccccccc}
\hline \hline
cluster & method A & \multicolumn{2}{c}{method B$^{1,2}$} && 
 \multicolumn{2}{c}{method B$^1$} & & method A$^3$ & method B$^3$ \\
annulus, & & & & & \multicolumn{2}{c}{fix $N_H$, use 0.5--2 keV} &
 & use all FOV & gain uncorrected \\
\cline{3-4} \cline{6-7}
arcmin & $T_e$, keV & $T_e$, keV &90\% & & $T_e$, keV & 90\% &
  & $T_e$, keV &$T_e$, keV \\
\hline
0--6   & 7.6 & 9.3 & 6.9--25  && 8.8 & 5.9--17  &&  8.5 & 6.1 \\
6--11  & 8.8 & 7.5 & 4.4--18  && 7.4 & 5.0--14  && 10.0 & 7.6 \\
11--18 & 3.9 & 3.3 & 2.2--6.2 && 3.8 & 2.6--6.2 &&  5.4 & 3.9 \\
18--25 & 2.1 & 1.8 & 1.2--11  && 1.7 & 1.1--7.3 &&  5.1 & 2.6 \\
\hline
\end{tabular}
\end{center}
$^1$ For these columns, the background error of 15\% and a systematic
flux error of 5\% are included. For other columns, only a background error of
5\% and no flux error are included.\\
$^2$ Temperatures from this column are shown in Fig.~1.\\
$^3$ These columns are included for comparison only; the values we consider
``correct'' are in the previous columns.
\vspace{-3mm}
\end{figure*}

\section{\rosat\ ANALYSIS OF A2256}

Six \rosat\ observations of A2256 --- a central exposure with the PSPC-C
detector (June 1990) and five offset pointings with the PSPC-B (low gain
state, October 1991--July 1992) --- were analyzed by Briel et al.\ (1991)
and BH. We reanalyze these data applying the updated calibration and a more
accurate technique to obtain temperatures in the image annuli same as those
used by M96 for \asca. Since we intend to compare our results to those of
BH, our analysis steps are given below with maximum detail. We accumulate
spectra from the annuli $r=0-6-11-18-25'$ centered on the main cluster
brightness peak. The regions containing the cooler infalling group (a
90\deg\ sector within $r=11'$) and the central galaxy (a 2\am\ radius
region) detected by BH, are excluded to facilitate comparison with the M96
results which are largely insensitive to the presence of such cool
substructure due to the higher energy band. Flat-fielded images in the
energy bands of 0.20--0.42--0.52--0.70--0.91--1.32--2.01 keV, free of
particle and most non-cosmic X-ray background, are generated using the
approach and code of Snowden et al.\ (1994). Cosmic (including any residual
non-cosmic) X-ray background is then calculated individually in each energy
band and for each pointing, using image regions more than 40\am\ from the
cluster center, within 28--50\am\ from the detector center and free of
detectable sources.  This results in a relative accuracy of the background
normalization of about 5\% (estimated by experimenting with different image
regions). We then employ and compare two techniques to correct for the
recently recognized SASS processing error (Snowden et al.\ 1995) which
introduces spurious gain variations over the detector face. This error has
especially serious implications for the results on clusters with
temperatures above $\sim 2.5$ keV, the \rosat\ upper energy cutoff. The
first method (method A hereafter) is to obtain cluster fluxes directly from
the flat-fielded images which are derived using the presently available
detector efficiency maps (Snowden et al.\ 1994) affected by the same error
in the same way, thus canceling the error to some extent (S. Snowden,
private comm.) The fluxes are then fitted using the on-axis spectral
response, binned in the broad energy bands. An advantage of this approach is
that it automatically includes all non azimuthally-symmetric variations of
the effective exposure, which are about $\pm$10\% on a several-arcminute
scale and are energy-dependent, as can be seen from comparison of the
exposure maps for different energy bands. These variations are due to a
combination of the spatial variations of the detector gain and efficiency. A
disadvantage of this method is that the detector efficiency maps in the wide
energy bands are, in fact, a convolution of the efficiency and the X-ray
background spectrum, which may be different from the object spectrum.
Another source of bias, which should be most prominent at large off-axis
angles, is a combination of the poor detector energy resolution and the
decline of the effective area with energy, such that the division of the
detector count rate by the effective area artificially increases the flux at
the highest energies --- a 1 keV photon which is detected as 2 keV is being
divided by the smaller effective area corresponding to 2 keV. Also, this
method is not optimal for minimizing statistical uncertainties, since it
assigns too high a weight to the outer regions with poorer statistics. It is
therefore preferable to use a technique which convolves a model with the
responses rather than attempting to deconvolve the data. In such a method
(hereafter, method B), we correct the SASS gain error for each event using
the code provided by the \rosat\ team, generate flat-fielded images using
S.~Snowden's code, calculate and subtract the sky background, then multiply
the background-subtracted images by the exposure maps (by which the raw
images were divided for flat-fielding) and fit fluxes from these images.
The model spectrum is multiplied by the azimuthally-symmetric off-axis area
weighted by the cluster surface brightness within each region, and convolved
with the binned photon redistribution matrix. The problems here are that at
present there are no detector maps corrected for the SASS gain error so the
background estimate may be less accurate, due to the spurious gain
differences between the outermost detector area used to calculate the
background and the central image area used for spectra (Snowden et al.\
1995).  Also, any azimuthally-asymmetric effective exposure variations that
are not due to the SASS gain error, are ignored. Since neither method is
fully satisfactory at present, we use both (with some additional
variations), and include the residual calibration uncertainties as
systematic errors. We will see that the accuracy of this first-order
approximation correction for the known problems is adequate for the purpose
of this paper.

The systematic errors are estimated as follows. The rms difference of the
background normalizations from the SASS gain error-corrected and uncorrected
data is about 15\%. We accept it as a reasonable estimate of the inaccuracy
due to the use of inconsistent detector efficiency maps, and assume a 15\%
error of the background in each energy band. This supersedes the 5\% error
due to the background calculation technique alone, mentioned above. The rms
difference of the cluster fluxes in our annuli between the corrected and
uncorrected data is about 5\%. Therefore, we chose to use a systematic error
in flux of 5\% in each of our energy bands as a reasonable representation of
the calibration uncertainties related to the use of the uncorrected detector
efficiency maps in method A and ignoring the spatial detector efficiency
non-uniformities in method B. This choice is somewhat arbitrary since the
errors of methods A and B are likely to be different, but is acceptable as
an order of magnitude estimate.  We do not include an error component due to
the presently uncorrected residual time-dependent PSPC gain variations
(Snowden et al.\ 1995), but our conservative 5\% flux error should encompass
that effect as well. Results obtained using different spectral response
matrices vary less than those from the different methods described here, so
we do not include the systematic errors due to the response matrix
uncertainties.

In all pointings, we use only the image area inside the detector support
structure ring, $\theta<18'$ to calculate the cluster fluxes, to minimize
the calibration uncertainties and the biases of method A. For comparison, we
have also performed the fitting using the data uncorrected for the SASS gain
error with symmetric off-axis areas (as in most earlier analyses), as well
as using the whole field of view. The free fitting parameters for each image
region were plasma temperature, Galactic absorption column and
normalization. Different pointings at the same region were fitted
simultaneously (without co-adding the data) using the response matrices
appropriate for the particular detector and its gain state. For the
outermost cluster annulus, two of the six exposures, both covering the same
South-West sector of the ring, give unacceptably high $\chi^2$, and we chose
to exclude them from the fit for this region. Including them does not change
the best-fit temperature significantly (changing it from 1.8 to 2.8 keV).
Results of all fits are listed in Table~1. The columns which we consider
``correct'' are those for methods A with $\theta<18'$ and B with the SASS
gain error corrected. Other two columns are included for comparison. We
chose to use method B as our preferred method to calculate temperatures in
the wide annuli for which the detector efficiency non-uniformities are
likely to average out. If $N_H$ is fixed at $5\times 10^{20}$ cm$^{-2}$ (the
value consistent with all regions) and only the upper four energy intervals
are fitted using this method, the temperatures remain unchanged, as is shown
in the table. Fig.~1 shows results from our preferred method with confidence
intervals calculated including only 5\% background error and no systematic
flux error (solid) and 15\% background plus 5\% flux errors (dashed
extensions), the best-fit values corresponding to the larger intervals. For
the brighter and hotter inner regions the flux uncertainty is most
important, while for the outer annulus it is the background uncertainty.
Separate fits to different pointings are shown in Fig.~1 {\em b} and {\em
c}.

Results from our ``correct'' methods, and even all but one best-fit value
from the ``incorrect'' methods in Table~1, are well within the 90\%
one-parameter confidence intervals shown for our preferred method. This
suggests that our systematic error estimates are adequate. This also
suggests that use of the presently unavailable self-consistent calibration
data to correct for the gain and detector efficiency variations, will not
qualitatively alter our results, since our two methods use a reasonable
approximation to correct for these effects. Calibration improvements will
reduce the values of the systematic errors that need to be included.

\begin{figure*}[tb]
{\centering\hbox to \textwidth{
\epsfysize=8cm
\epsffile{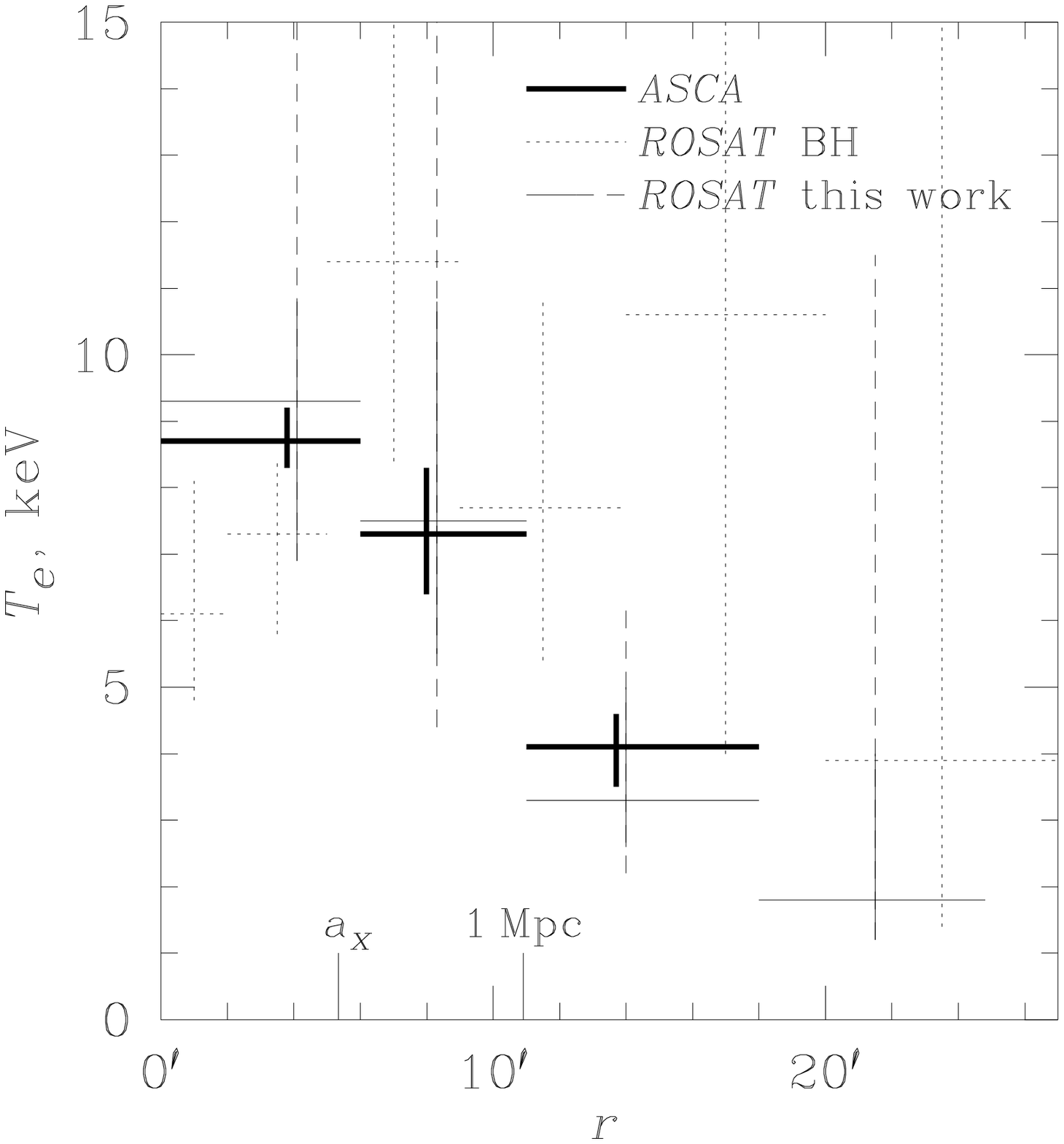} \hfill
\epsfysize=8cm
\epsffile{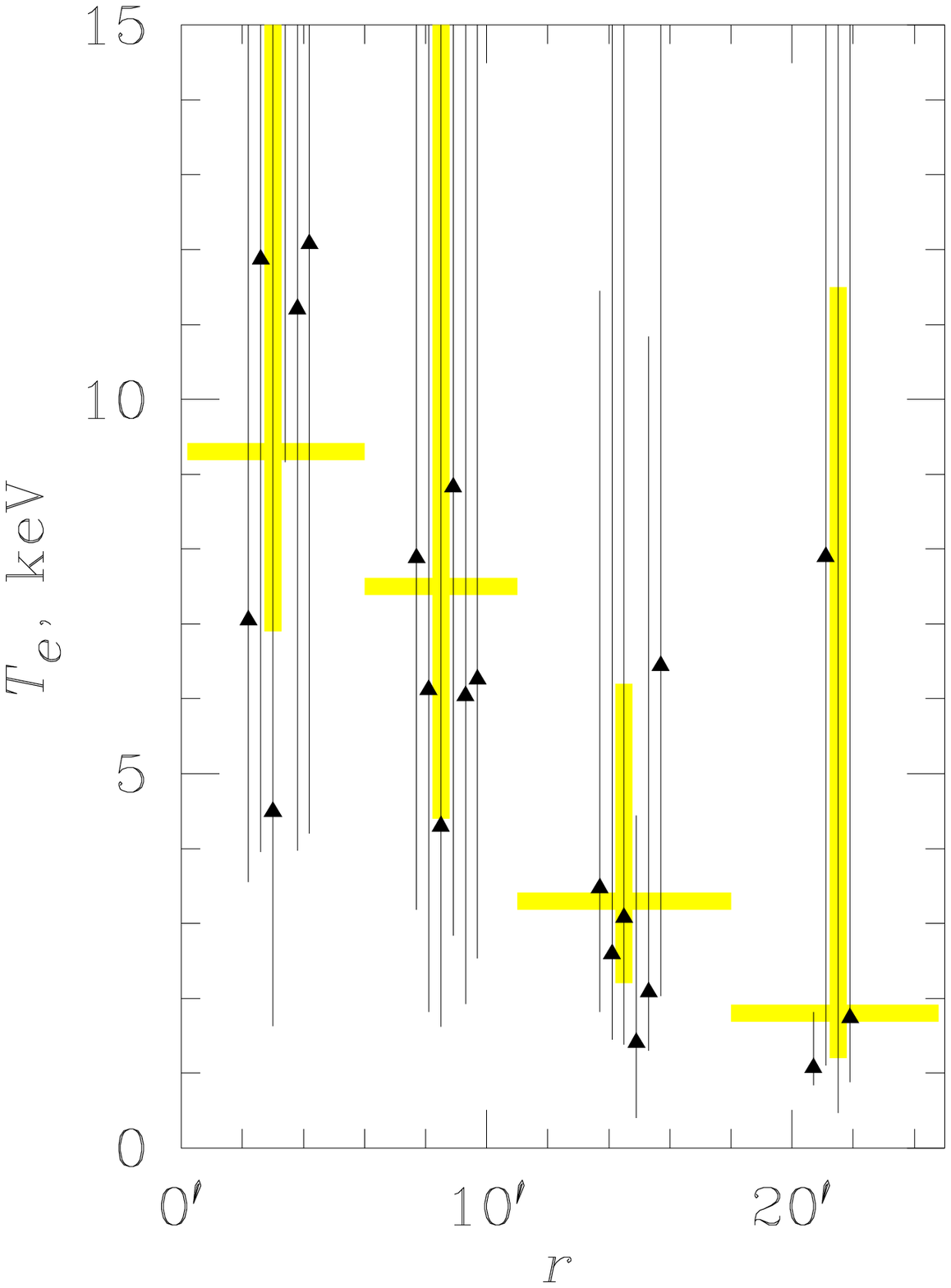} \hfill
\epsfysize=8cm
\epsffile{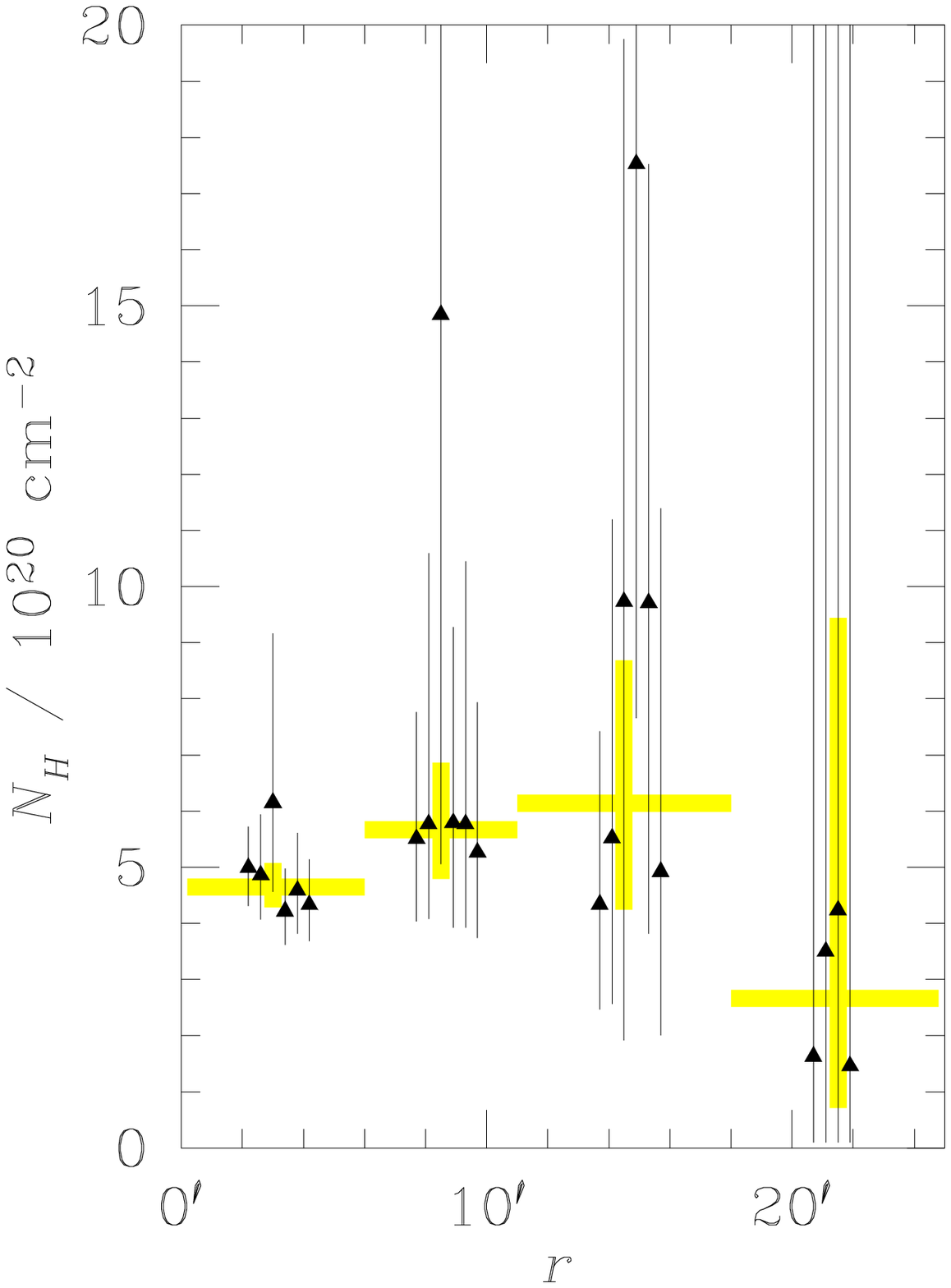}
}}
\vspace{1mm}
\small\parindent=3.5mm
{\sc Fig.}~1.---({\em a}) Projected temperature profiles of A2256 from
\asca\ (Markevitch 1996) and \rosat\ (BH and J. P. Henry, private comm.;
this work) with 90\% confidence intervals (BH's 68\% intervals are
multiplied by 1.65). \asca\ and our \rosat\ analysis exclude from the first
and second annuli cooler regions of the infalling group and (for \rosat) the
central cluster galaxy, found by BH. Values from BH are for the entire
annuli, thus they are not directly comparable to ours in the inner 10\am.
Dashed extensions of our \rosat\ error bars show the effect of inclusion of
the systematic errors.  ({\em b,c}) Values of temperature and hydrogen
column density for individual \rosat\ pointings in our analysis for the same
regions as in panel ({\em a}). Gray lines show values for all pointings
fitted simultaneously, as in panel ({\em a}). There is no significant radial
trend in $N_H$, even without inclusion of the systematic errors (such a
trend, if existed, might indicate deficiencies of our analysis procedure),
while there is a trend for the temperature to decrease with radius, in
agreement with \asca.
\end{figure*}

\subsection{Comparison with earlier ROSAT results}

Fig.~1 shows that our \rosat\ and \asca\ temperature profiles are in
excellent agreement, and \rosat\ results even suggest that the temperature
decrease found by \asca\ continues at still larger radii. However, earlier
\rosat\ results obtained by BH from the same data are different (although
only at about a 90\% significance). We summarize the differences between BH
and our analysis techniques which may be responsible for the discrepancy, in
addition to the correction of the SASS spatial gain error.

1. Use by BH of the whole field of view and division of each count by a
vignetting factor. The problems of this approach, included in the discussion
of our method A above, are spurious enhancement of the high-energy flux at
large off-axis angles and assignment of too much weight to low-statistics
data. As Table~1 shows, inclusion of the outer detector area noticeably
increases the best-fit temperatures, which may also be partially due to the
less accurate effective area calibration there.

2. Gain adjustment. BH adjusted the PSPC gain in each pointing so that their
overall cluster temperature becomes 7.5 keV, the value obtained by \ginga.
We note that in the presence of cool substructure, this should introduce an
error. Indeed, assuming after Briel et al.\ (1991) that the cool subgroup
has $T_e=2$ keV and contributes about 1/6 of the flux, while the rest of
cluster has $T_e=8$ keV, \ginga\ should derive a 7.2 keV single-temperature
fit while \rosat\ would find 5.7 keV because of the different spectral
coverage. We have not performed any gain adjustments but excluded the group
region from our analysis.

3. Use of photons up to 2.5 keV. Our analysis is restricted to $E<2.0$ keV.
According to the \rosat\ User's Guide, the calibration is not reliable above
that energy, with the expected discrepancy between the model and data
reaching 20\%.

4. Summation of data from pointings which are performed with different
detectors. We have fitted different pointings simultaneously but without
co-adding the data, making use of the information on the detector identity
and their differing properties.

We have also fitted the temperatures in the regions of the hot spots found
by BH which lie 7\am\ to the North-East and South-West of the cluster
center. Using method A to account for the non-uniformities of the detector
efficiency, and including data within $\theta=18'$ from the detector center, we
obtain temperatures of 8.8 ($>$4.4) keV (90\% interval) and 11.1 ($>$5.3)
keV for the more significant NE spot and less significant SW spot,
respectively, if $N_H$ is free, and 14.6 ($>$6.9) keV and 4.8 (3.2--8.2) keV
if it is fixed. No systematic errors were included except for the 5\%
background error, and inclusion of the flux error makes confidence intervals
even larger. Our other methods give different best-fit values but never
significantly different from the cluster mean. The SASS gain error
correction reduces the temperature of the NE spot and increases that of the
SW spot. Inclusion of $\theta>18'$ data (which for these regions means
offset pointings) changes the best-fit temperatures to 12.5 (5.6--24) keV
and 14.5 (6.9--23) keV for the NE and SW spots, respectively, if $N_H$ is a
free parameter, and to 18.3 (8.7--24) keV and 6.0 (4.2--9.2) keV if it is
fixed.  We note, however, that these spots fall on the PSPC detector support
ring in three (four) offset pointings out of five for the NE (SW) spots. We
conclude that the hot spots reported by BH are probably artifacts.

\begin{figure*}[ht]
\pspicture(0,5.6)(18.8,20)

\rput[tl]{0}(-0.2,20.1){\epsfxsize=9cm
\epsffile{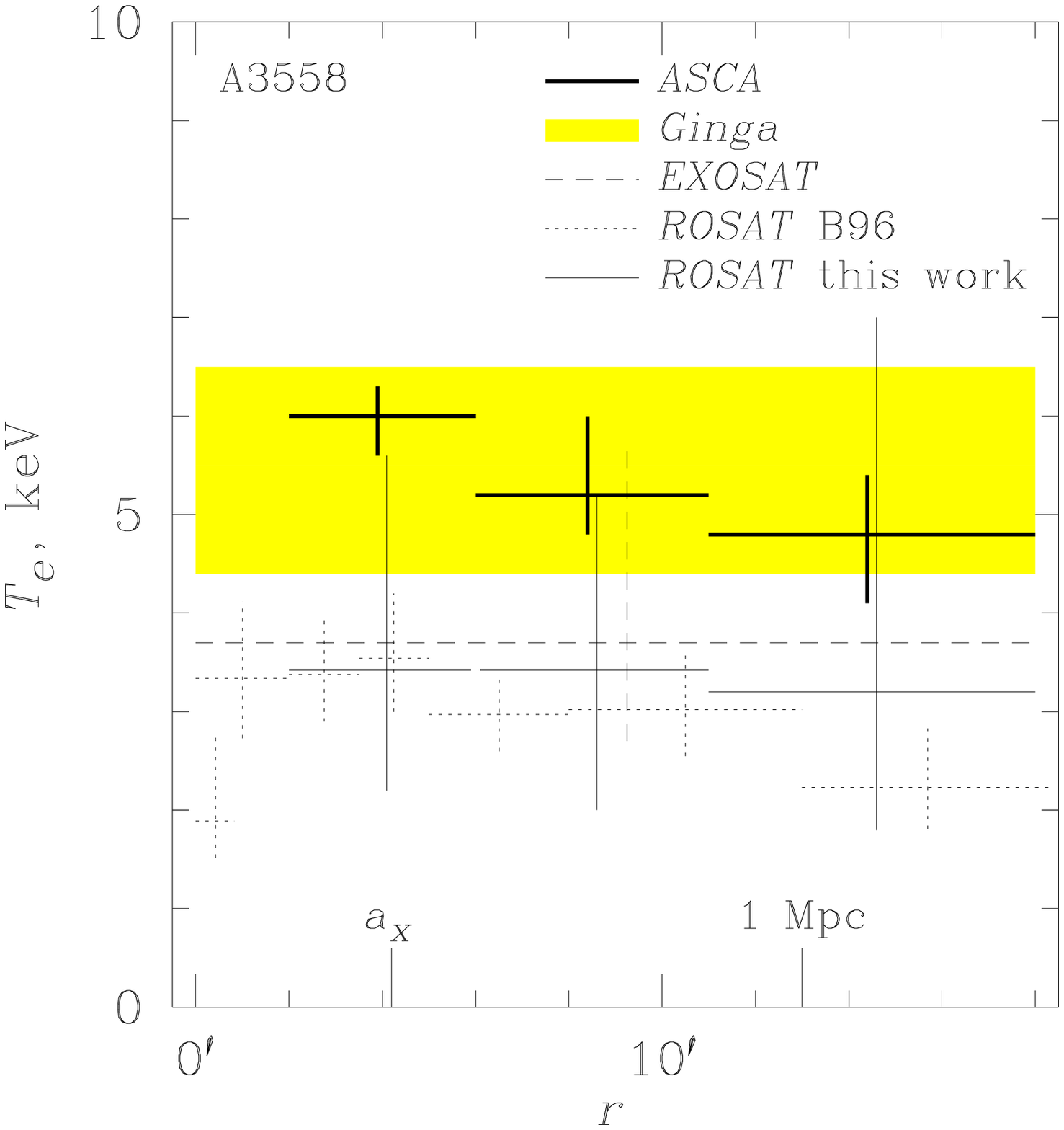}}

\rput[tl]{0}(9.3,20.1){\epsfxsize=9cm
\epsffile{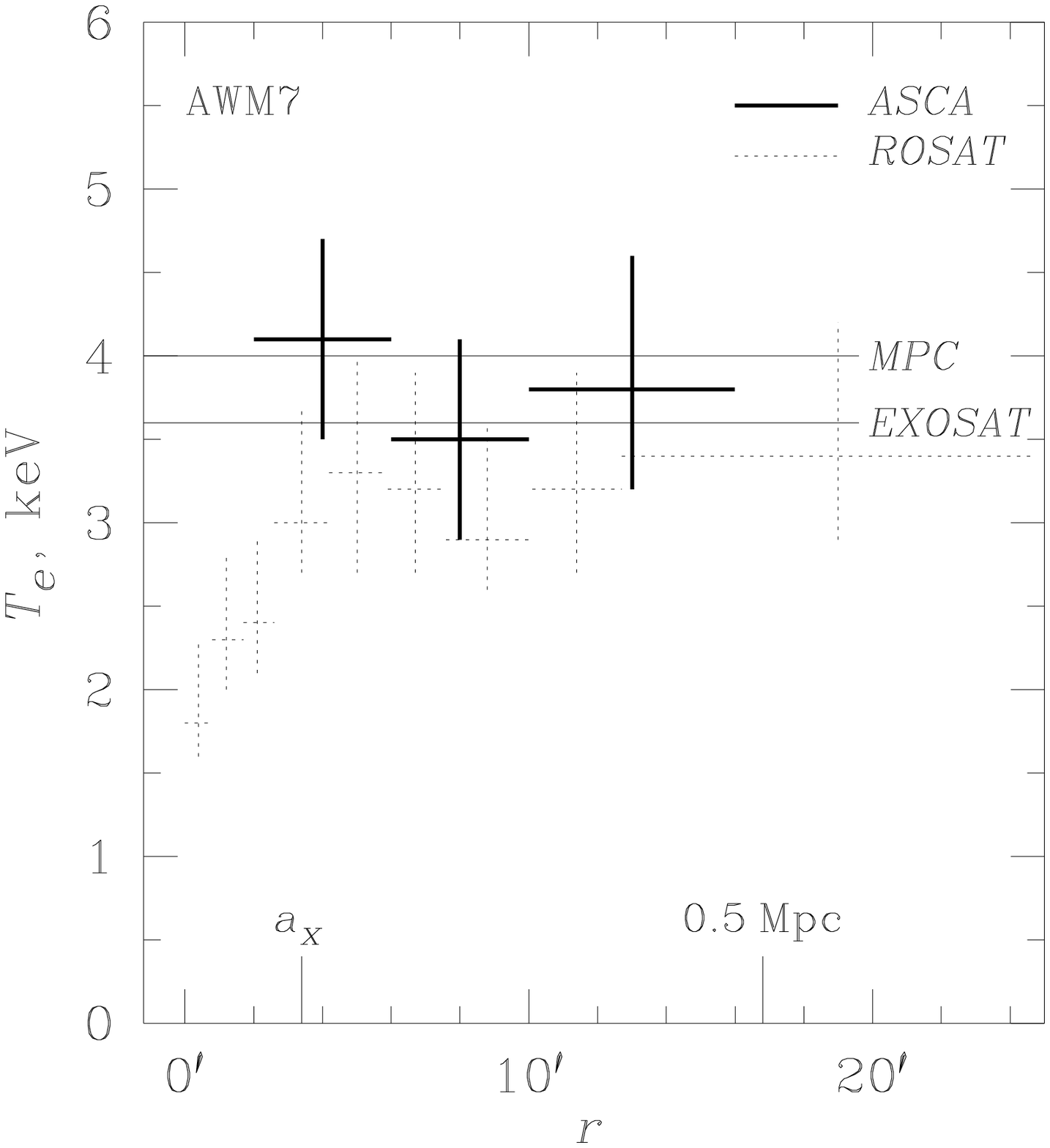}}

\rput[tl]{0}(-0.1,10.3){
\begin{minipage}{9cm}
\small\parindent=3.5mm
{\sc Fig.}~3.---Projected temperature profiles of A3558 from \asca\ (this
work), \rosat\ (B96 and this work), and the overall temperature from
\ginga\ (Day et al.\ 1991) and \exosat\ (Edge \& Stewart 1991). Errors are
90\% confidence. The gray band shows the temperature range of the \ginga\
main spectral component for different models considered by Day et al. \asca\
annuli are centered as shown in Fig.~2 (while they are narrower than those
in Fig.~2). The core radius is estimated excluding the cD galaxy (B96). The
\asca\ single-temperature fit to the central region is 5.6 keV; however, its
spectrum requires more than one component (see the text) and is not directly
comparable to the \rosat\ single-temperature fit. Our \rosat\ values are
similar to those from B96 when no systematic errors are included. Inclusion
of systematic errors (shown) considerably enlarges the confidence intervals.
\end{minipage}
}

\rput[tl]{0}(9.7,10.3){
\begin{minipage}{9cm}
\small\parindent=3.5mm
{\sc Fig.}~4.---Projected temperature profiles of AWM7 from \asca\ (this
work), \rosat\ (NB) and the overall temperature from the \einstein\ IPC
(David et al.\ 1993) and from \exosat\ (Edge \& Stewart 1991). Errors are
90\% confidence.  Annuli are centered on the cD galaxy. The core radius is
from an estimate excluding the cD galaxy (NB). The \asca\ temperature value
for the cD galaxy region (3.9 keV single-temperature fit) is discussed in
the text. \rosat\ values outside the cD region are systematically lower than
those from other instruments.
\end{minipage}
}
\endpspicture
\end{figure*}

\section{A3558}

\subsection{ASCA results}

We restrict our present discussion to a comparison of results from
\asca\ and other instruments; a scientific interpretation will be presented
elsewhere. We have analyzed the GIS and SIS data simultaneously to maximize
statistical significance of the results, even though the SIS $22'\times 22'$
field of view does not cover the whole cluster. The useful exposure for
A3558 is 17 ks for the GIS and 12 ks for the SIS (which was operated partly
in the 4-CCD mode and partly in the 2-CCD mode). To obtain a two-dimensional
temperature map and a radial temperature profile of this cluster, we
followed the technique of Markevitch et al.\ (1996) and M96 with some
modifications regarding the central region. This technique consists of
simultaneous fitting of the temperature in all image regions, taking into
account the \asca\ energy-dependent mirror scattering. The mirror PSF was
modeled using Cyg~X-1 data for energies above 2~keV (Takahashi et al.\
1995). Unlike the clusters considered in M96, A3558, as well as AWM7, have a
central dominant galaxy and a cooler component in the center, known e.g.,
from the \rosat\ data (below). Also, average temperatures of these clusters
are lower.  Therefore, it is desirable to include energies below 2.5 keV,
which were excluded by M96 due to the PSF and other uncertainties. In the
present analysis, we use energies starting from 1.5~keV (excluding the
poorly calibrated 2--2.5~keV interval). For energies below 2~keV, the PSF
data from the 2--3 keV interval were used after the appropriate correction
for the energy dependence of the GIS spatial resolution. Comparison of this
extrapolation of the PSF model to the brightness profile of a point source
as in Takahashi et al.\ shows that its inaccuracy is less than 10\% for the
interesting range of radii. We use this value as a relative PSF uncertainty
in the annuli, while assuming a 15\% uncertainty for sectors during the
temperature map reconstruction. As in M96, we use a \rosat\ PSPC image as
the surface brightness template, correcting it for the non-isothermality
under the assumption that the projected temperature is constant over each
region, with the exception of the region around the cD galaxy. For these
central regions with a possibly complex spectrum, this technique is
inadequate and the model relative normalization (set by the \rosat\ image
for other regions) is fitted as a free parameter together with the
temperatures.

B96 who analyzed \rosat\ data on A3558, found that the cD galaxy peak is
displaced by 1.3\am\ with respect to the centroid of the surrounding
cluster. We centered our inner region on the cD galaxy but all other rings
on the cluster centroid, both for the map and the profile analysis. Our
resulting temperature map overlaid on the \rosat\ image is shown in Fig.~2.
Sectors of the annuli in Fig.~2 were chosen to sample the cluster regions
approximately parallel and perpendicular to the possible merger direction
(see the image). Projected temperatures in the annuli with $r=2-6-11-18'$
are shown in Fig.~3 (such angular resolution is at the limit of \asca\
capabilities at the present calibration accuracy). There is a nearby poor
cluster $\sim 25'$ to the East of A3558 and another one still farther in the
same direction (Breen et al.\ 1994; B96). A sector containing part of the
nearby cluster was included in the analysis to account for the scattering of
its flux especially into our region 9 (Fig.~2), although its contribution
was found to be negligible. Its temperature was fixed at 4 keV. The stray
light contribution from these sources is negligible because of their low
flux and the favorable value of the telescope roll angle used in this
observation (Ishisaki 1996).

In the cD galaxy region, a single-temperature fit results in $T_e=5.6$ keV
and $\chi^2=270$ with 256--14 d.o.f.\ (for the simultaneous fit of the
spectra from the annuli). The fit is significantly improved if an additional
spectral component is allowed in the central region. For a simple model
consisting of two thermal components, we obtain the
temperatures\footnote{Errors are 90\% single-parameter confidence intervals
throughout the paper. \asca\ errors are estimated by simulations including
systematic uncertainties.} of $1.1^{+0.6}_{-0.3}$ keV and 12 ($>8$) keV with
approximately equal projected emission measure in the $r=2'$ cD galaxy
region (this region has about a quarter of the total cluster emission
measure), and $\chi^2=240$ with 256--16 d.o.f. This model is consistent with
the \rosat\ PSPC single-temperature fit to the same region. A power-law
second component fits the data equally well ($\chi^2$=245) but the resulting
photon index is unphysical (+0.5). A power law with the index fixed at the
typical AGN value of --1.7 improves the fit less significantly. A more
detailed analysis and discussion will be presented elsewhere; here we only
note that there definitely is some hard spectral component in the central
region, because there is excess flux in that region at the highest \asca\
energies.

Using different models for the central region does not change the results
for the rest of the cluster considerably. There is a significant asymmetric
temperature pattern in the map (Fig.~2), with the North-East sector of the
cluster hotter than other regions. The temperature values in regions 2-9
could not be due to statistical deviations from the same mean at the 99.9\%
confidence; the temperatures in regions 2-5 of the second annulus are
different at the 98\% confidence. The results of fitting SIS and GIS data
separately are in good agreement, and this temperature pattern is present in
both fits. There is also a slight radial decline of the cluster temperature
(Fig.~3), although it is marginally significant. We obtained an average
cluster temperature of $5.5^{+0.3}_{-0.2}$ keV.

\subsection{Comparison with other instruments}

A3558 was observed by \exosat\ (Edge \& Stewart 1991) in an offset pointing
from which a poorly constrained temperature of $3.8^{+2.0}_{-1.0}$ keV was
obtained. Day et al.\ (1991) reported a \ginga\ overall temperature of
$5.7\pm0.2$ keV for this cluster and found that the spectrum requires more
than one component. Additional components they considered were a power law
with $\gamma=-1.7$ or a cooler gas, which both significantly improved the
fit. \asca\ results presented above are in good agreement with these
measurements. \rosat\ PSPC data were analyzed by B96 who obtained a
significantly lower temperature of $3.3^{+0.4}_{-0.3}$ keV for the inner
$r=5'$ excluding the cD galaxy and a slight decrease of the temperature
outward. B96 discussed their inconsistency with \ginga\ and suggested that
the \ginga\ wide-aperture flux may be contaminated by emission from nearby
clusters. We note that, although the \ginga\ field of view included the
nearest clusters, they are far less luminous and are expected to be cooler
as well. Therefore, they are unlikely to produce such a temperature
overestimate. We have reanalyzed the \rosat\ data and obtained results
almost identical to those from B96, using our method B without the SASS gain
error correction and not including the systematic errors to facilitate
comparison. The SASS error correction did not change the temperature values
noticeably for this observation. However, if the systematic errors like
those we used above for A2256 are included, the error bars are increased
sufficiently so that the results become consistent with \asca\ and \ginga\
within the 90\% confidence intervals. Indeed, to raise the PSPC-derived
temperatures to the \asca\ values, only a 7--8\% increase in the flux is
necessary in our highest PSPC energy band. Our \rosat\ values for the same
outer three annuli used in the \asca\ analysis, including the systematic
errors, are shown in Fig.~3. Note that the systematic errors are not of
statistical nature and may not necessarily be reduced by averaging over the
entire cluster. They will be reduced by improvements in the calibration. One
of such calibration-related errors is a possible anomalous PSPC gain in this
particular observation (performed in July 1991 with PSPC-B in high gain
state) which may drive all best-fit temperatures down. A gain change of 3\%
would suffice to produce the observed discrepancy. It is unlikely that any
correction for this effect would drastically change the radial temperature
distribution. Except for the absolute values, the \asca\ and \rosat\
projected temperature profiles in Fig.~3 both are consistent with a slightly
decreasing temperature, outside the central region. We note that the
anomalously high gas to total mass fraction in this cluster, obtained by B96
using the \rosat\ temperatures for the total mass calculation, may be
overestimated by a factor of about 1.7 because of this temperature
uncertainty.

\section{AWM7}

This nearby poor cluster was observed by \asca\ with several offset
pointings. The complete dataset is analyzed by Ezawa et al.\ (1996), who
included a detailed treatment of stray light contamination which is
important for offset pointings. Earlier, the central pointing was analyzed
by Ohashi et al.\ (1994). These authors found that the gas temperature in
this cluster does not change with radius but a cooler component and an
abundance increase is required in the central region containing the cD
galaxy. Here we limit our analysis to only the central pointing for which
the stray light is unimportant, and derive a two-dimensional temperature
distribution and a radial temperature profile out to 16\am\ from the cluster
center, using the same method as in Section 3. For this cluster, only GIS
data are used for simplicity, which is sufficient for the technical purpose
of this paper. For the $r=0-2-7-16'$ annuli centered on the cD galaxy and
the outer 2 rings divided into 4 (or 5) sectors, we detect no significant
temperature deviations from the cluster average of $3.9\pm0.2$ keV. \asca\
temperatures in the narrower, $r=2-6-10-16'$ concentric annuli are presented
in Fig.~4.  Since the cluster emission is present beyond our outermost
annulus, the \rosat\ emission to $r=20'$ from the cluster center is included
in the model for the last ring (both for the map and the profile analysis)
to avoid boundary effects. Our temperature profile is consistent with an
isothermal one, similarly to the Ezawa et al.\ results in the same range of
radii. The single-temperature models for all regions are acceptable; in the
central $r=2'$ region, our single temperature is $3.9\pm0.7$ keV. However,
if lower energies starting from 0.7 keV are included in our analysis (only
for this estimate), the addition of a cooler component in the region of the
cD galaxy improves the fit. For two thermal components, the best-fit
temperatures are $4.8^{+1.8}_{-0.9}$ keV and $1.3\pm0.3$ keV with the cool
component contributing a fraction of the projected emission measure within
$r=2'$ equal to $0.24\pm0.12$. This model is consistent with the \rosat\
single temperature fit in the same region.

Fig.~4 also shows results from two other high energy but spatially
unresolved observations of this cluster, with the \einstein\ MPC (David et
al.\ 1994) and \exosat\ (Edge \& Stewart 1991). \asca\ agrees well with both
instruments. Also shown in the figure is the \rosat\ PSPC temperature
profile obtained by NB from the observation performed in January 1992
(PSPC-B low gain state). This profile lies systematically below all three
high-energy measurements even outside the cooler cD galaxy region, although
the temperatures in individual radial bins are consistent with them within
the 90\% confidence intervals. Including only the 5\% systematic background
error, we obtained an average \rosat\ temperature of $3.1\pm0.2$ keV for the
cluster excluding the $r=3'$ central region and the area outside
$\theta=18'$. This value is similar to the median value of the temperature
profile from NB but is significantly below the values from other
instruments. However, the discrepancy is less prominent than that for A3558
discussed above. The same systematic error as that used above, is certainly
sufficient to account for the difference. Both \asca\ and \rosat\ profiles
agree in that there is no significant temperature change with radius outside
the cD galaxy region. Note that for this nearby galaxy group, our
measurements cover a much smaller linear size at the group's distance than
for other clusters studied.

\section{SUMMARY}

We have presented a reanalysis of the \rosat\ PSPC observations of A2256 and
of the \asca\ central pointing of AWM7, as well as first \asca\ results on
A3558. Our primary goal has been to compare \asca\ cluster temperatures with
those from other instruments; thus, we have focused on the technical issues
of data analysis rather than physical interpretation. We find that:

1. The \rosat\ data on A2256 are in excellent agreement with recent
\asca\ findings (M96), including the detection of the temperature decline
with radius, although with much larger errors than those of \asca. Hot
gas regions earlier reported by BH from the \rosat\ data are probably
artifacts;

2. The \rosat\ and \asca\ temperature profiles for A3558 and AWM7
qualitatively agree. For AWM7, there is no significant temperature
variations with radius outside the cD galaxy region. For A3558, there is a
slight radial temperature decline outside the central region.  For the cD
galaxy regions, the instruments agree when a multi-temperature spectrum is
allowed;

3. \asca\ values of the overall temperature for A3558 and AWM7 (5.5 and 3.9
keV, respectively) are in agreement with those derived by other high-energy
instruments. However, the \rosat\ absolute temperature values for these
clusters are lower. We note that, although formally statistically
significant, the discrepancy is within the current \rosat\ calibration
uncertainties, which we propose as its likely cause.

These findings add confidence in the \asca\ results on other clusters, for
which \asca\ is the only instrument capable of mapping the spatial
temperature distribution. They also underline the high sensitivity of the
\rosat\ temperature measurements for hot clusters to the calibration
uncertainties. We also find that:

4. \asca\ two-dimensional temperature map of AWM7 (within $r<16'$) does not
display any significant spatial temperature variations outside of the cD
galaxy regions. An asymmetric temperature pattern is found in A3558;

5. From \asca\ data, the central region of A3558 contains a source of
emission harder than the cluster average and perhaps harder than the typical
AGN power law emission, in addition to a cooler component.

\acknowledgments

This work was motivated by discussions at the ``X-ray Imaging and
Spectroscopy'' conference held in Tokyo in March 1996. We thank Steve
Snowden for a useful discussion regarding the \rosat\ calibration, and Bill
Forman and Craig Sarazin for valuable comments on the manuscript. MM was
supported by NASA grants NAG5-2526 and NAG5-1891. AV received support from
the Smithsonian Institution.

\end{multicols}

\vspace{1cm}
\pspicture(0,3.8)(18.8,20)

\rput[tl]{0}(-0.2,20.2){\epsfxsize=9.5cm
\epsffile{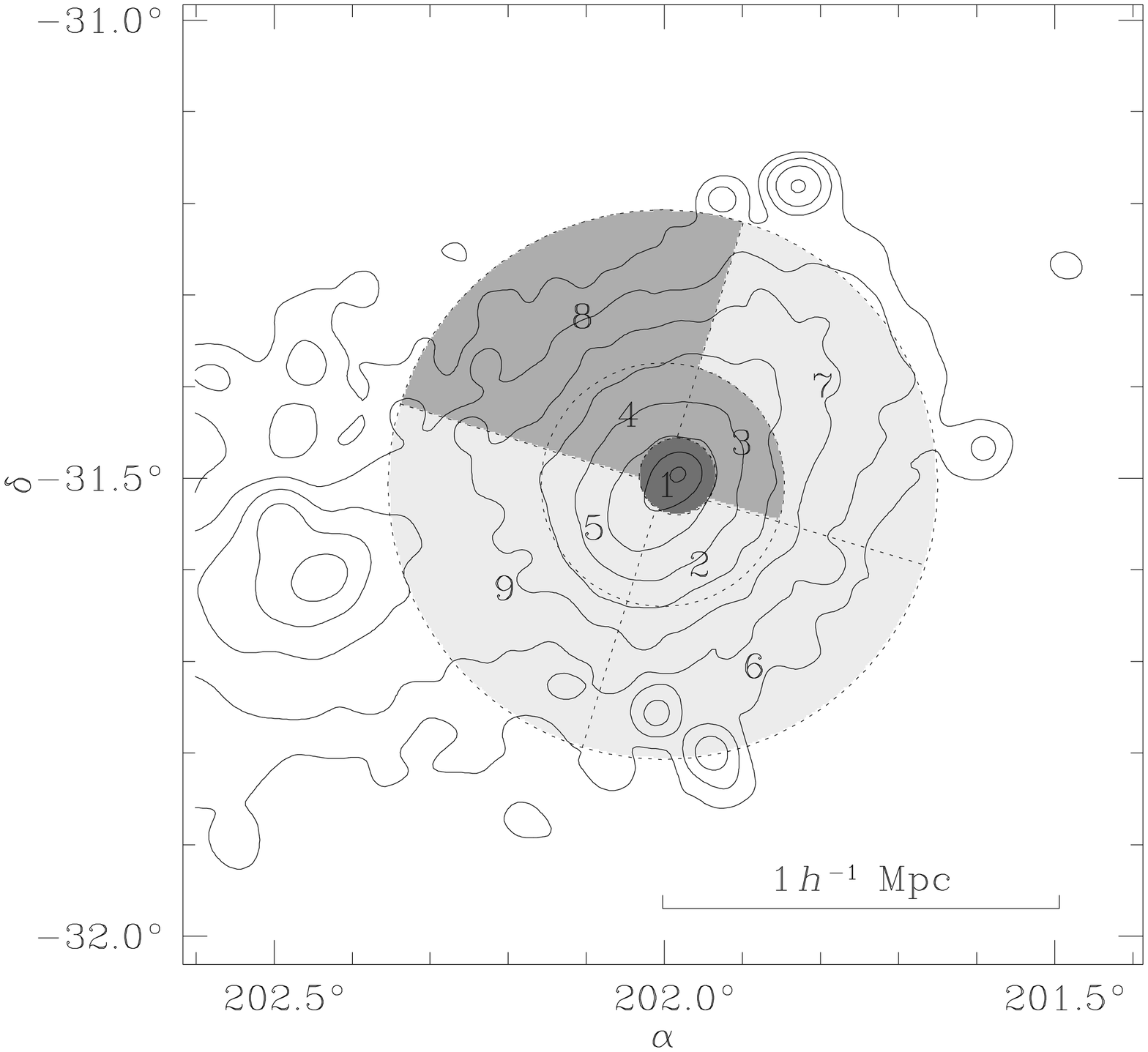}}

\rput[tl]{0}(0.32,11.9){\epsfxsize=8.77cm
\epsffile{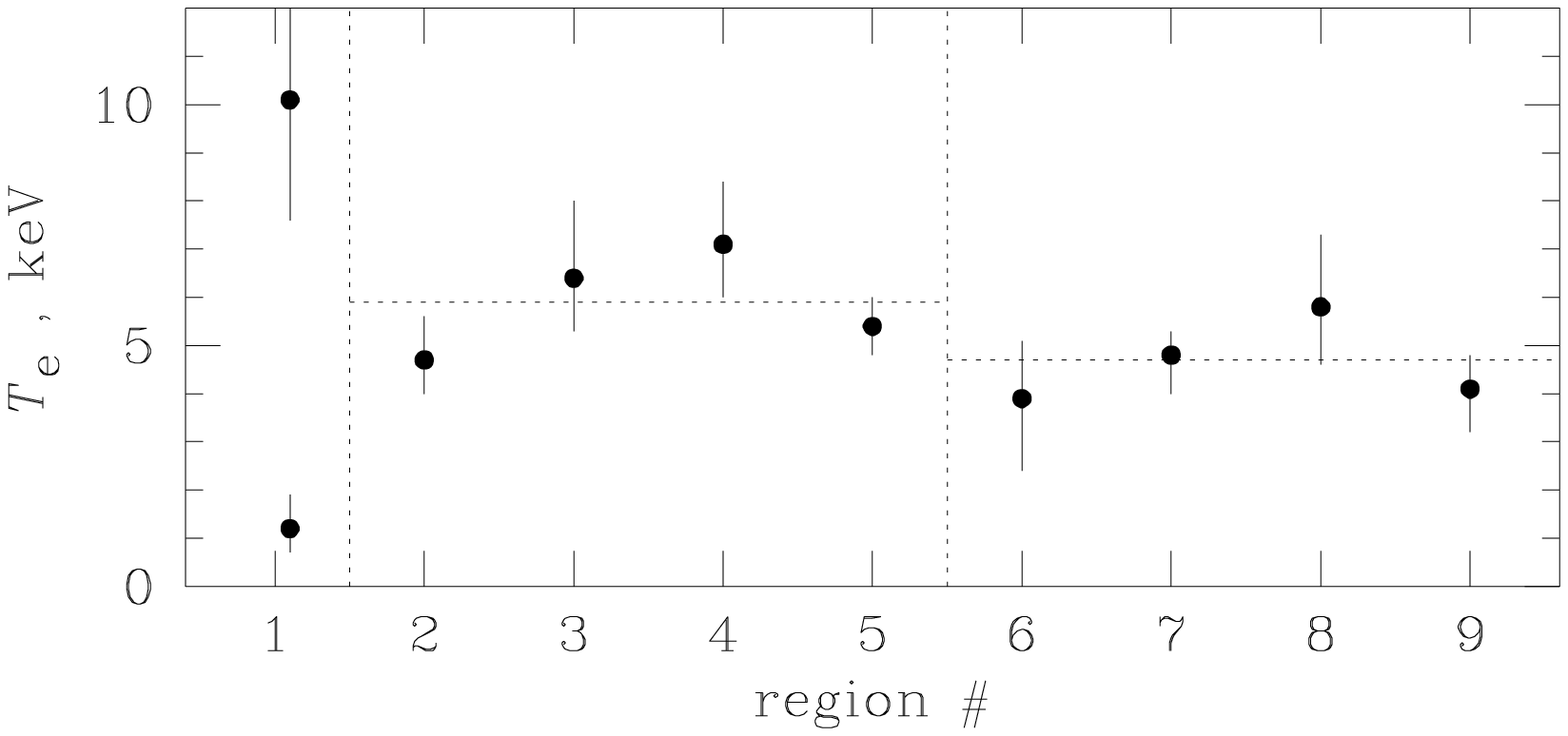}}

\rput[tl]{0}(-0.1,7.3){
\begin{minipage}{9cm}
\small\parindent=3.5mm
{\sc Fig.}~2.---\asca\ temperature map of A3558. Contours are \rosat\
surface brightness and dashed lines are boundaries of the regions in which
the \asca\ temperature is reconstructed. The inner $r=2.5'$ circle is
centered on the cD galaxy while other rings (8\am\ and 18\am) are centered
on the cluster centroid determined by B96. The temperatures are shown by
grayscale (hotter is darker). Regions are numbered and their temperature
values with 90\% confidence intervals are shown in the lower panel. For the
central region, the result of a two-temperature fit is shown (see the text),
and the grayscale arbitrarily corresponds to the higher-temperature
component. Dashed horizontal lines in the lower panel show average
temperatures in each ring.
\end{minipage}
}
\endpspicture


\begin{references}

\reference{} Bardelli, S., Zucca, E., Malizia, A., Zamorani, G., Scaramella,
R., \& Vettolani, G. 1996, A\&A, 305, 435

\reference{} Breen, J., Raychaudhury, S., Forman, W., \& Jones, C. 1994,
ApJ, 424, 59

\reference{} Briel, U. G., et al.\ 1991, A\&A, 246, L10

\reference{} Briel, U. G., \& Henry, J. P. 1994, Nature, 372, 439 (BH)

\reference{} Churazov, E., Gilfanov, M., Forman, W., \& Jones, C. 1996,
in ``X-ray Imaging and Spectroscopy of Cosmic Plasmas'', ed.\ F. Makino, in
press

\reference{} Day, C. S. R., Fabian, A. C., Edge, A. C., Raychaudhury, S.
1991, MNRAS 252, 394

\reference{} David, L., Slyz, A., Jones, C., Forman, W., Vrtilek, S. D., \&
Armaud, K. A. 1993, ApJ, 412, 479

\reference{} Edge, A. C., \& Stewart, G. C. 1991, MNRAS, 252, 414

\reference{} Ezawa, H., Fukazawa, Y., Haiguang, X., Kikuchi, K., Makishima,
K., Ohashi, T., Tamura, T., Yamasaki, N. 1996, in ``X-ray Imaging and
Spectroscopy of Cosmic Plasmas'',  ed.\ F. Makino, in press

\reference{} Ishisaki, Y. 1996, PhD thesis, University of Tokyo

\reference{} Loewenstein, M., 1995, presented by R. Mushotzky in
``R\"{o}ntgenstrahlung from the Universe'', eds.\ H. U. Zimmermann et al.,
MPE Report 263

\reference{} Loewenstein, M., Mushotzky, R., \& Donahue, M. 1996, in
``X-ray Imaging and Spectroscopy of Cosmic Plasmas'', ed.\ F. Makino, in
press

\reference{} Markevitch, M., Mushotzky, R. F., Inoue, H., Yamashita, K.,
Furuzawa, A., \& Tawara, Y. 1996, ApJ, 456, 437

\reference{} Markevitch, M. 1996, ApJ, 465, L1 (M96)

\reference{} Neumann, D. M., \& \bohringer, H. 1995, A\&A, 301, 865

\reference{} Ohashi, T., et al.\ 1994, in ``New Horizon of X-ray
Astronomy'', eds.\ F. Makino \& T. Ohashi (Tokyo: Universal Academy)

\reference{} Snowden, S. L., McCammon, D., Burrows, D. N., Mendenhall, J. A.
1994, ApJ, 424, 714

\reference{} Snowden, S. L., Turner, T. J., George, I. M., and Yusaf, R.
1995, OGIP Calibration Memo CAL/ROS/95-003

\reference{} Takahashi, T., Markevitch, M., Fukazawa, Y., Ikebe, Y., Ishisaki,
Y., Kikuchi, K., Makishima, K., \& Tawara, Y. 1995, \asca\ Newsletter, \# 3
(NASA/GSFC)

\reference{} Tanaka, Y., Inoue, H., Holt, S. S. 1994, PASJ 46, L37

\reference{} White, S. D. M., Navarro, J. F., Evrard, A. E., \& Frenk, C. S.
1993, Nature, 366, 429

\end{references}
\end{document}